\RequirePackage[2020-02-02]{latexrelease}
\documentclass[times, twoside]{zHenriquesLab-StyleBioRxiv}

\leadauthor{Henriques} 

\begin{document}

\title{Harnessing Artificial Intelligence \\ To Reduce Phototoxicity in Live Imaging}
\shorttitle{AI-enabled Live Imaging}

\author[1,*]{Estibaliz Gómez-de-Mariscal}
\author[1,*]{Mario Del Rosario}
\author[2]{Joanna W Pylvänäinen}
\author[2,3,4]{Guillaume Jacquemet}
\author[1,5\Letter]{Ricardo Henriques}
\affil[*]{Equally contributed authors}
\affil[1]{Optical cell biology group, Instituto Gulbenkian de Ciência, Oerias, Portugal}
\affil[2]{Faculty of Science and Engineering, Cell Biology, Åbo Akademi University, Turku, Finland}
\affil[3]{Turku Bioimaging, University of Turku and Åbo Akademi University, Turku, Finland}
\affil[4]{InFLAMES Research Flagship Center, Åbo Akademi University}
\affil[5]{MRC Laboratory for Molecular Cell Biology, University College London, London, United Kingdom}

\maketitle

\begin{abstract}

Fluorescence microscopy, widely used in the study of living cells, tissues, and organisms, often faces the challenge of photodamage. This is primarily caused by the interaction between light and biochemical components during the imaging process, leading to compromised accuracy and reliability of biological results. Methods necessitating extended high-intensity illumination, such as super-resolution microscopy or thick sample imaging, are particularly susceptible to this issue. As part of the solution to these problems, advanced imaging approaches involving artificial intelligence (AI) have been developed. Here we underscore the necessity of establishing constraints to maintain light-induced damage at levels that permit cells to sustain their live behaviour. From this perspective, data-driven live-cell imaging bears significant potential in aiding the development of AI-enhanced photodamage-aware microscopy. These technologies could streamline precise observations of natural biological dynamics while minimising phototoxicity risks.

\end{abstract}


\begin{corrauthor}
rjhenriques\at igc.gulbenkian.pt
\end{corrauthor}

\section*{Introduction}

The ability to comprehend biological dynamics is inherently linked to the capacity for non-invasive observation. Fluorescence microscopy has been instrumental in facilitating these analyses across a range of scales, with molecular specificity~\cite{heimstadt_fluoreszenzmikroskop_1911, reichert_fluoreszenzmikroskop_1911, lehmann_luminszenz-mikroskop_1913}. Over the past two decades, technological advancements such as light sheet microscopy~\cite{huisken_optical_2004, dodt_ultramicroscopy_2007, verveer_high-resolution_2007, reynaud_lsm_perspective_2008}, structured illumination microscopy (SIM)~\cite{heintzmann_super-resolution_2017, gustafsson_surpassing_2000}, and single molecule localisation microscopy (SMLM)~\cite{betzig_imaging_2006, hess_fpalm_2006, lelek_smlm_2021}, have revolutionised fluorescence light microscopy, enabling us to characterise biological events from molecular interactions up to larger living organisms.

\begin{figure} 
\centering
\includegraphics[width=1\linewidth]{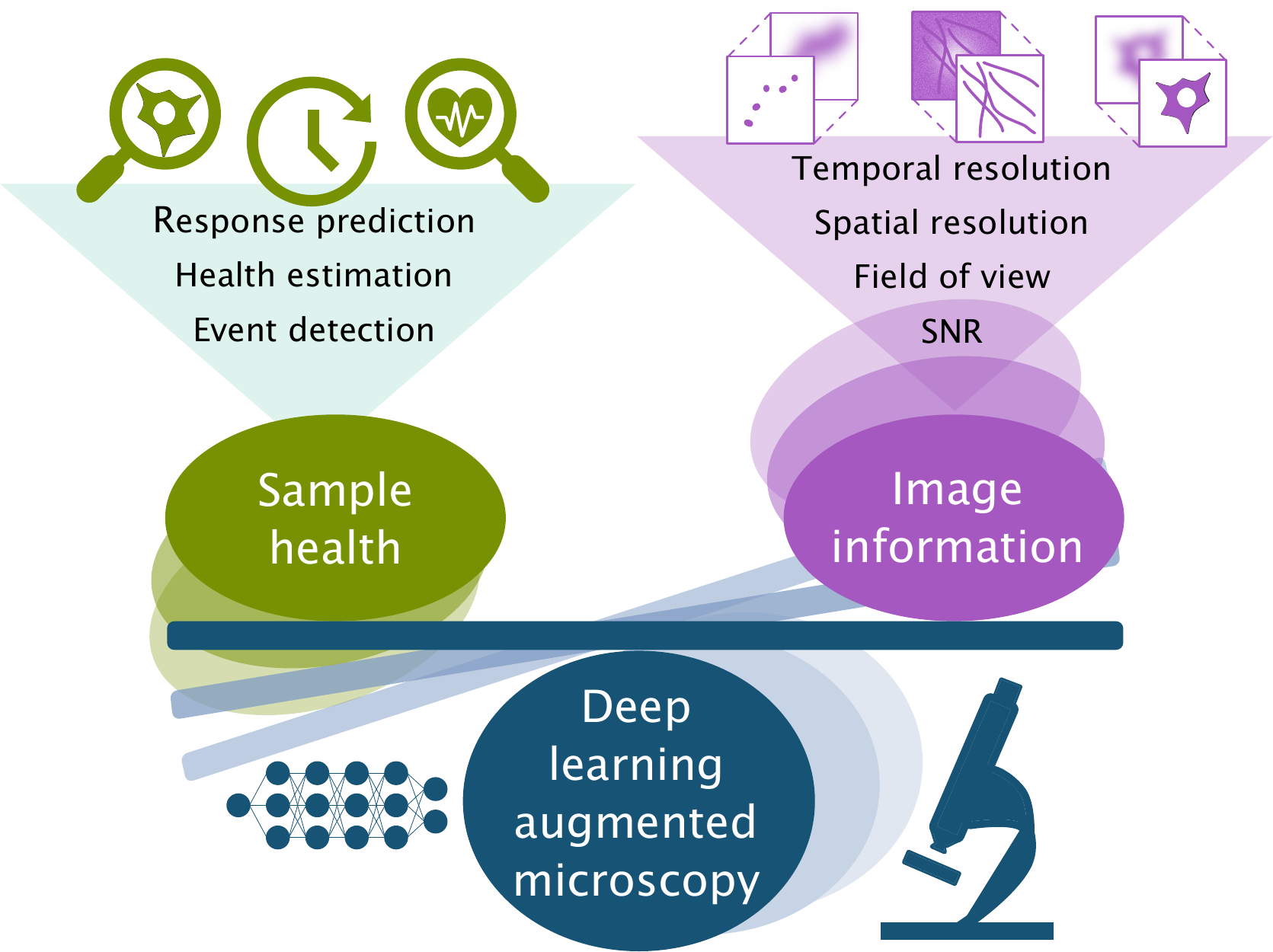}
\caption{Graphical abstract. The delicate balance between sample health and the information obtained by imaging requires a compromise between both elements. Deep learning augmented microscopy aims to reduce this compromise, striving to obtain equal information from our sample with less impact on its health. SNR: Signal-to-noise ratio}
\label{fig:graphical}
\end{figure}

A key challenge associated with advanced microscopy analysis is its likely need to increase fluorescence excitation light levels, which results in phototoxicity or photodamage. These terms refer to the detrimental impacts of light, especially when employing photosensitising agents or high-intensity illumination~\cite{waldchen_light-induced_2015, tinevez_quantitative_2012, laissue_assessing_2017}. Despite certain distinctions like photodamage occurring also in non-living materials, both terms are here used interchangeably, for simplicity. Sample illumination may also result in photobleaching, a process characterised by an irreversible loss of a fluorescent signal attributed to the destruction of the fluorophore. This is one manifestation of light damage, among other possible effects. Phototoxicity severely influences the experimental outcomes by altering biological processes under observation, skewing findings and impeding consistency. Therefore, it's crucial during live-cell microscopy to carefully consider these factors to prolong imaging durations and achieve dependable research outcomes.

The biological validity of live-cell imaging experiments requires a precise balance between the data quality and the specimen's health, as depicted in Fig.~$1$. Major advancements have been made in both hardware and software technologies aiming to reduce sample light damage. Hardware innovations such as Lattice Light Sheet Microscopy (LLSM)~\cite{chen_lattice_2014} and Airyscan Microscopy~\cite{huff_airyscan_2015} are notable examples. Additionally, computational advancements like Fluctuation-based Super Resolution Microscopy offer promising solutions to these issues~\cite{dertinger_fast_2009, gustafsson_fast_2016}. A recent study has shown that a two-colour illumination scheme combining near-infrared illumination with fluorescence excitation presents an interesting capacity to limit the phototoxicity caused by light-induced interactions with fluorescent proteins~\cite{ludvikova_natbiotech_2023}. These technological breakthroughs have the potential to optimise observation accuracy while mitigating photodamage.

In parallel, artificial intelligence (AI), specifically deep learning, can significantly improve imaging information in low illumination scenarios by considerably enhancing image quality and quantification~\cite{belthangady_applications_2019, tian_deep_2021, melanthota-2022}. This has inspired the search for integrated solutions by the microscopy community~\cite{wagner-natmeth-2021, ebrahimi-commbio-2023, bouchard_tagan_machineintelligence_2023}. The fusion of advanced optical hardware with computational models and intelligent analytics heralds new breakthroughs in overcoming sample damage induced by traditional live fluorescence microscopy methodologies, marking the advent of AI-enhanced intelligent microscopy.

\begin{figure*} 
\centering
\includegraphics[width=\linewidth]{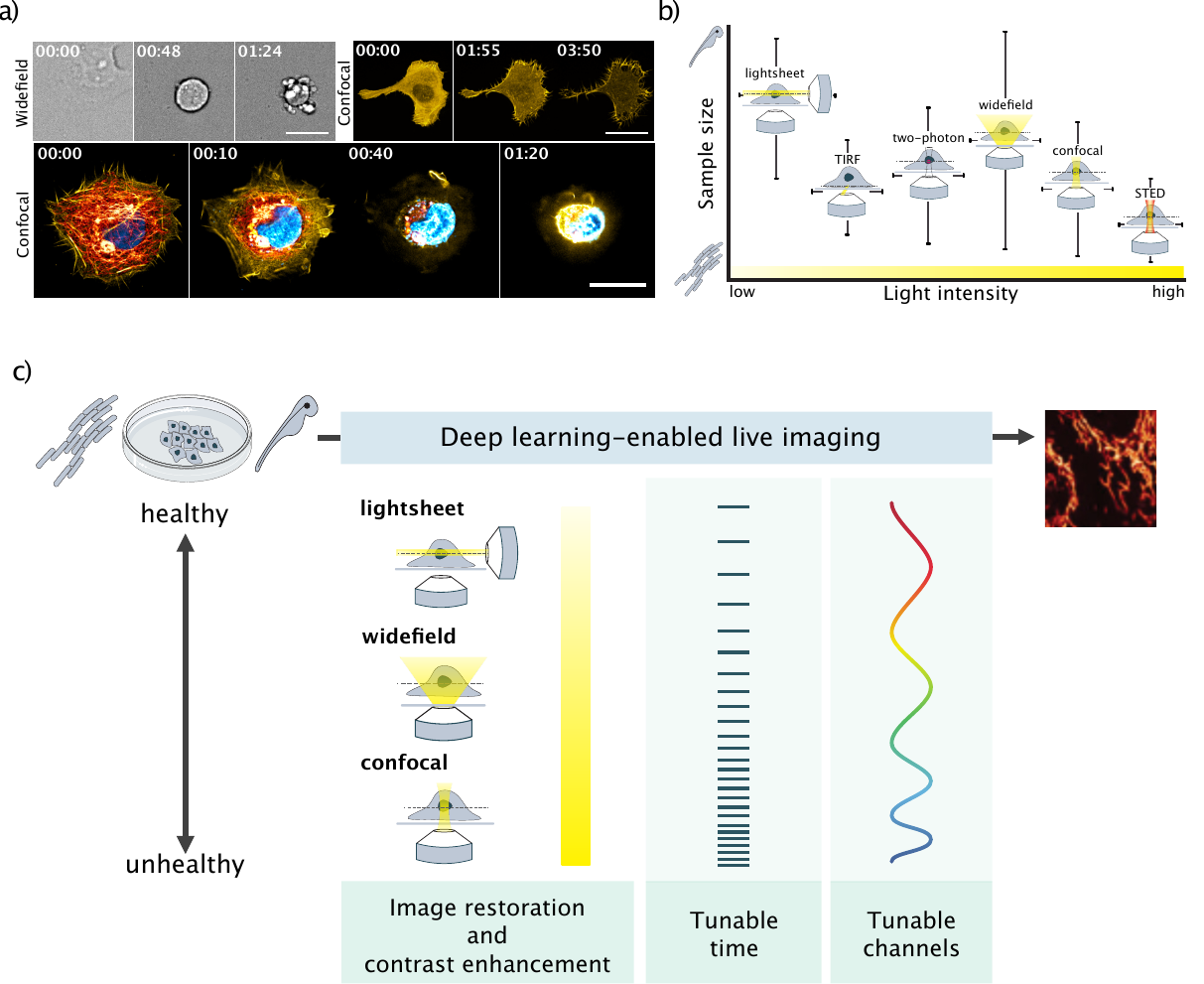}
\caption{\textbf{Phototoxicity and sample health during live-cell acquisition.} a) Examples of photodamage across widefield and confocal microscopy. Cellular rounding up and membrane blebbing are observed in widefield; actin depolymerisation and cellular collapse are seen in confocal. yellow: actin; orange: tubulin; cyan: nuclei. b) The effect of different light microscopy modalities according to the type of illumination on the imaged sample. Some modalities, such as Lightsheet microscopy, decrease the amount of light on the sample to increase cell survival. Other modalities sacrifice sample health to gain spatial resolution, such as STED c) AI-enabled low phototoxicity live-cell microscopy by reducing the amount of light needed on the sample to obtain the same information. A few examples are shown, such as image restoration and contrast enhancement applicable to different microscopy modalities, acquisition sampling reduction by tuning time acquisitions, and fluorescent channel tuning. The scale bar represents $25\mu m$.}

\label{fig:1}
\end{figure*}

\section*{Understanding phototoxicity}

Cellular environments undergo continuous adaptation to maintain homeostasis, with oxygen playing a key role in multiple chemical processes~\cite{sies_reactive_2020}. Reactive Oxygen Species (ROS), resulting from oxidative processes involving oxygen radicals (Table~1), are integral to cell growth, differentiation, and other functions. However, uncontrolled ROS processes lead to oxidative stress and cellular damage, resulting in autophagy, inflammation, and potentially cell death~\cite{sies_reactive_2020}. For these reasons, a balance between these states is critical for cellular physiology.\\

\begin{table}[htbp]
    \centering
    \caption{\textbf{Oxygen radicals and examples of biomolecules affected by oxidative processes}. The most common oxygen radicals present during ROS processes~\cite{sies_reactive_2020} and biomolecules affected across different levels, interfering with molecular pathways and cellular signalling~\cite{eichler_flavins_2005, hockberger_activation_1999, fraikin_role_1996, ricchelli_photophysical_1995, cunningham_photosensitized_1985}. }
    \label{tab:oxygen_radicals}
    \begin{tabular}{ll}
        \hline
        \textbf{Oxygen Radicals} & \textbf{Biomolecules Affected} \\
        \hline
        Hydrogen Peroxide & Flavins \\
        Singlet molecular oxygen & Porphyrins \\
        Nitric Oxide & NAD(P)H \\
        Superoxide anion radicals & Tyrosine \\
        Hydroxyl radical & Catecholamines \\
        Hydroxide ion & Cysteinyl thiols \\
        \hline
\end{tabular}
\end{table}

Light irradiation to excite fluorescence, in addition to inducing oxidative stress by producing high amounts of ROS, triggers multiscale alterations across biological samples that alter the homeostasis of oxidative processes. Moreover, higher doses of light irradiation increase this oxidative stress, further aggravating cellular homeostasis. Photochemical processes at the molecular level lead to intracellular component damage or toxic compound formation~\cite{laissue_phototoxicity_2021, tosheva_between_2020}. The literature extensively documents the impacts of fluorescence excitation light, especially UV light on DNA, such as DNA double-stranded breaks~\cite{zhang_phototoxic_2022}, thymidine dimerisations~\cite{zhang_phototoxic_2022}, UV-induced apoptosis~\cite{mateos-pujante_evaluation_2022}, and tumour factor activation~\cite{mateos-pujante_evaluation_2022}. These effects are even routinely employed in research settings under controlled conditions~\cite{10.1269/jrr.10175, shibai_mutation_2017, pfeifer_mutations_2005, kumar_uv_2015, tan_selection_2021}. Moreover, molecules naturally present in cells (Table~$1$) undergo degradation via exposure to light-induced oxidative stress, canonically produced during the fluorescence excitation process. This degradation culminates in the generation of reactive oxygen compounds that directly impair cell health by oxidising DNA, proteins, unsaturated fatty acids, and enzyme cofactors \cite{laissue_assessing_2017, icha_phototoxicity_2017}. Excessive ROS shifts the homeostasis of the cell cytoplasmic environment' leading to organelle damage like mitochondrial fragmentation due to compromised membrane potential and integrity, among others~\cite{alam_characterization_2022, zhang_phototoxic_2022, kim_phototoxicity_2015, mcdonald_light-induced_2012, waldchen_light-induced_2015}. 

While helpful and versatile, fluorescence microscopy illumination entails an additional source of ROS formation~\cite{demchenko_photobleaching_2020}. When excited, fluorophores undergo autocatalysis through dioxygen, releasing hydrogen peroxide and consequently degrading - a process commonly known as photobleaching~\cite{nienhaus_chromophore_2016}. Photobleaching produces ROS similar to other biomolecules, intensifying phototoxic effects~\cite{demchenko_photobleaching_2020}. However, despite the interrelation between photobleaching and phototoxicity~\cite{ludvikova_natbiotech_2023}, these phenomena exhibit distinct features and can occur independently. Namely, identical oxygen radical compounds originate from photobleaching and direct fluorescence excitation light interactions with other cellular components~\cite{demchenko_photobleaching_2020, nienhaus_chromophore_2016}. Suggesting that a reduction of photobleaching does not necessarily imply a decrease in phototoxicity, and the other way around. The intricate mechanisms underlying these phenomena have been thoroughly researched, especially concerning various fluorophores' photosensitising effects in live microscopy. Given the common necessity for oxygen in cell cultures, ROS formation during live-cell imaging is an unavoidable aspect of the imaging process.

At lower thresholds, the impact of light exposure may be minimal and reversible, allowing the cell to return to physiological conditions. This cell recovery will depend on several factors, namely how resilient the sample is and the degree of phototoxicity incurred by the experiment \cite{waldchen_light-induced_2015}. In severe cases, the behaviour of cells will be permanently altered~\cite{laissue_assessing_2017}. Moreover, further alterations normally accumulate within a cell's homeostatic environment until reaching an irreversible point causing pronounced impairment of cell health that leads to unwanted cell behaviour (cellular collapse) or death (apoptosis) (Figure~$2$a) \cite{icha_phototoxicity_2017, laissue_assessing_2017, tosheva_between_2020, zhang_phototoxic_2022}.

Although scientists have identified several indicators of photodamage, the correlation between fluorescent excitation light dose and light damage is not yet fully understood due to variability in experimental conditions. These variations include culture conditions, illumination modes (Figure~$2$b) and biological sample variability, hindering the understanding between fluorescent excitation light dose and light damage~\cite{laissue_assessing_2017, tinevez_quantitative_2012}. Despite challenges associated with varying specimen resistances to light damage, establishing universal quantitative benchmarks across diverse samples could harmonise these effects and enhance reproducibility in the imaging context.

\section*{Phototoxicity quantification}

The literature describes several phototoxicity hallmarks with numerous known markers available to identify and characterise sample damage~\cite{laissue_phototoxicity_2021}. However, using these markers presents additional challenges in live-cell experiments. Firstly, incorporating phototoxicity markers necessitates extra planning and may require allocating fluorescent channels usually reserved for observing conditions of interest (\emph{e.g.}, markers for DNA oxidative damage). Not only does this limit the number of fluorescent channels available for the experiment, but when paired with live cell experiments could enhance ROS formation, thereby escalating the phototoxicity risk on the specimen. Secondly, despite well-documented phototoxicity markers, quantification-based screenings are less commonly employed~\cite{tosheva_between_2020}. The absence of live-cell imaging universal metrics relatable to light exposure and consequent damage across various biological systems hampers experimental reproducibility, undermines results' robustness, and limits the obtainable biological readouts. Furthermore, without these universal metrics, it's challenging to fully leverage the capacity to image biological systems. This results in the incorrect assessment of experimental conditions to achieve maximum spatial and temporal resolution while preserving cell viability.

Researchers primarily rely on observation and experience to assess cell health and viability during photodamage evaluation (Figure~$2$a)~\cite{waldchen_light-induced_2015, tosheva_between_2020, laissue_assessing_2017}. While there are some attempts to provide quantifiable metrics for improving sample viability~\cite{tinevez_quantitative_2012, icha_phototoxicity_2017, waldchen_light-induced_2015}, they often simplify the impact of fluorescence excitation light to a binary classification of viable/healthy or non-viable/dead. These approaches may overlook subtle effects that could disturb a specimen's normal physiology. A gradient model that considers the accumulation of such discrete minor effects would more accurately depict the spectrum of effects documented in the existing literature. Namely, these approaches could be even more flexible by considering both cell health decline and recovery.

Yet an important aspect to consider is the collective physiological and behavioural outcome, rather than individual effects. In controlled experimental conditions, specimens typically exhibit consistent behaviours such as cell division, motility, and membrane dynamics. By modelling and quantifying these behaviours, deviations can be correlated to the negative effects of fluorescent light exposure on the sample.

Phototoxicity is a known issue in live-cell imaging, and a plethora of strategies exist to mitigate its effects, ranging from reducing light irradiation by either reducing the acquisition points or the light dose \cite{reynaud_lsm_perspective_2008} to using more sensitive light detectors \cite{huff_airyscan_2015}. Other strategies focus on controlling oxidative stress effects in biological samples by reducing its effects by supplementing antioxidants \cite{harada_antioxidant_2022,kesari_plant-derived_2020}, or increasing the resistance of the sample itself \cite{kunkel_increasing_2018}. Although some strategies are simpler to integrate than others, they require using specific and costly equipment or altering the conditions of the specimen. 

As advanced image analysis tools become increasingly available, a future strategy for fully reproducible live-cell imaging procedures could incorporate specific phototoxicity reporters within automated image acquisition and analysis workflows. This approach, paired with standardised experimental guidelines for identifying and quantifying phototoxic events, represents a promising solution. As imaging technologies evolve, adopting such methods should be prioritised by scientists aiming to create robust imaging strategies for visualising biological phenomena.

\section*{The relationship between fluorescence excitation light, image information and phototoxicity}

Fluorescence live-cell imaging involves the excitation of fluorophores using light, resulting in photon emission. These photons can be captured by a camera or sensor, facilitating the visualisation of cellular targets. Various microscopy methods are available for sample illumination as depicted in Figure~$2$b. Some strategies, such as widefield, confocal microscopy, and Stimulated Emission Depletion (STED), illuminate the sample across the optical axis. In contrast, two-photon excitation primarily excites sample regions near the imaging focal plane while also subjecting the sample to less harmful infrared light across the optical axis. Total Internal Reflection Fluorescence (TIRF) imaging optically limits illumination to near the coverslip surface, substantially reducing in-depth illumination. However, it generally only illuminates the interface between the sample and the coverslip. Light-sheet microscopy, a more live-cell-friendly approach, illuminates only around the focal plane of the sample but often poses challenges in achieving high resolution compared to alternative methods. Hence, the choice of microscopy depends on sample characteristics; more sensitive specimens like embryos benefit from gentler modalities such as light-sheet microscopy. Inversely, more resilient samples can withstand intense irradiation experienced during Single-Molecule Localization Microscopy (SMLM) or STED microscopy (Figure~$2$b). 

As introduced earlier, modern microscopy technologies seek to minimise the required illumination by becoming more specific towards the type of information that needs to be visualised. However, the acquired information's quality may depend on factors such as the signal-to-noise ratio (SNR), image contrast, and temporal and spatial resolution of the image data (Figure~$2$c). Each microscopy equipment has inherent limitations that affect these properties' optimisation, necessitating a trade-off, which is often expressed by the microscope 'pyramid of frustration'~\cite{weigert_content-aware_2018}, or as characterised explicitly for super-resolution microscopy by Jaquemet et al.~\cite{jacquemet_cell_2020}. This trade-off can be optimised based on experimental needs by adjusting light exposure intensity and acquisition speed - common parameters used to achieve less harmful imaging configurations. As such, methods that enhance image quality from sample-friendly setups, like deep learning methods, are particularly valuable when pushing against intertwined image information and phototoxicity limitations (Figures~$1$ and~$2$c). Specifically, deep learning's growing ability to augment and refine image-based information is attracting considerable interest in the microscopy community. It is becoming one of the most popular strategies to enable imaging setups with reduced phototoxicity~\cite{Scherf_gentlemic_2015, ebrahimi-commbio-2023}. In succeeding sections, we explore different strategies conceived to enhance microscopy image data's information contrast and quality, which could facilitate varied live-cell imaging setups with decreased fluorescent light illumination.

\section*{Deep learning for microscopy to the rescue}

\begin{figure*}
\centering
\includegraphics[width=1\linewidth]{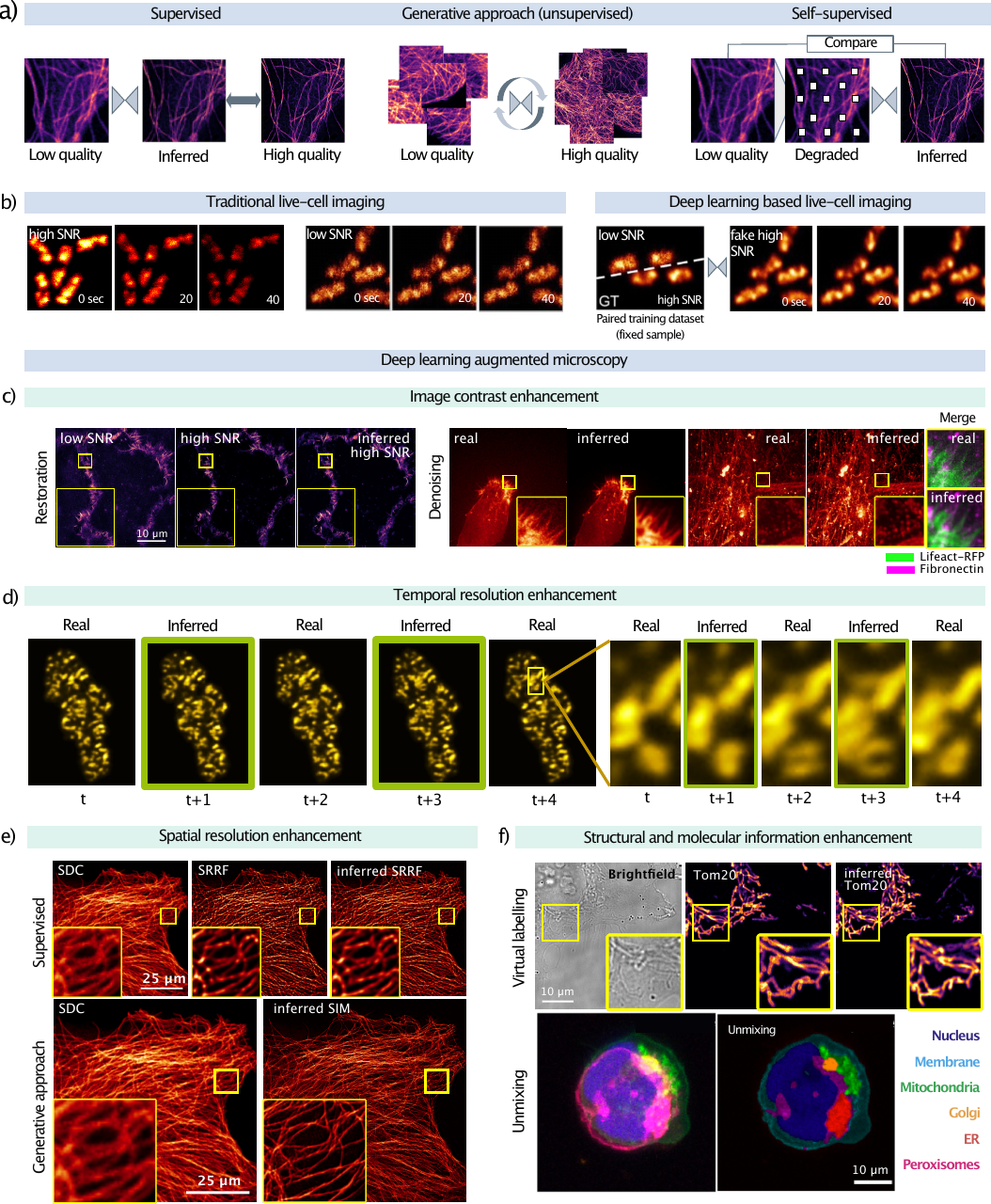}
\caption{\textbf{The deep learning landscape for a gentler live-cell microscopy imaging}. a) Types of deep learning methods according to the training strategy. b) High illumination intensities are used to obtain images with high SNR at the expense of causing photobleaching, among others. Reducing the fluorescence illumination intensity prevents photobleaching at the expense of obtaining images with low SNR. These common limitations in traditional live-cell imaging can be overcome by using deep learning to enhance the image contrast and extend the acquisition in a gentler manner. Additionally, one could image fixed samples to create the training data. c-f) Deep learning models can be used to run microscopy acquisitions that use lower fluorescence light intensities or illuminate the sample less often. For this, after the imaging experiment, one could c) restore and denoise images; d) improve the temporal resolution by inferring intermediate time points; or e) virtually super-resolve images with supervised super-resolution or with generative approaches for cross-modality style transfer. f) One could avoid using fluorescence illumination with virtual staining or partially, by using unmixing approaches that can decouple different structures from autofluorescence or crosstalk between channels. a), c), e) were extracted and modified from~\cite{von_chamier_democratising_2021}; b) and d) from~\cite{spahn_deepbacs_2022}; and f) from~\cite{von_chamier_democratising_2021} for the virtual labelling and~\cite{mcrae_plosone_2019} for the unmixing. SNR: signal-to-noise ratio.  \\}
\label{fig:figure3}
\end{figure*}

The exceptional ability of deep learning to automatically discern, uncover, and summarise complex patterns within images is undeniably one of its most transformative contributions to bioimaging. It has significantly influenced microscopy image analysis by ushering in a shift from traditional mathematical feature modelling to data-driven methods~\cite{Moen2019, belthangady_applications_2019, meijering_birds-eye_2020, tian_deep_2021, melanthota-2022}. We have observed the potential of deep learning to achieve remarkable accuracy and, in some cases, human-level performance in diverse computer vision tasks such as segmentation, denoising, detection, reconstruction, unmixing, and response prediction. Additionally, emerging imaging processing tasks like cross-modal style transfer~\cite{wang_deep_2019, hollandi_nucleaizer_2020} and virtual labelling~\cite{christiansen_silico_2018, ounkomol_label-free_2018-1, xu_gan-based_2019, rivenson_virtual_2019} hold promising prospects due to the flexibility they introduce when designing imaging experiments (\emph{e.g.,} spectral unmixing or rapid low-resolution acquisitions that can be virtually super-resolved for subsequent quantification)~\cite{belthangady_applications_2019}. Indeed, microscopy imaging naturally allows for creating paired image datasets by alternating acquisition setups and combining different modalities~\cite{qiao_evaluation_2021, von_chamier_democratising_2021}, simulating data~\cite{sage-natmeth2019, nehme_deep-storm_2018, fang_deep_2021, saguy-dblinknatmeth-2023, hyun-Jic-diffusion-2023}, or, recently, developing correlative approaches such as CLEM~\cite{boer-clem-2015}. Cumulatively, all these advancements have laid a solid foundation for the growing field of deep learning-augmented microscopy (Figure~$3$).

Regardless of the strategy, the key idea behind deep learning is to define models that learn to identify or enhance specific features directly from the data. Traditionally, the learning process, or training, has been classified as supervised, where the model is trained with paired input-output image datasets, or unsupervised, where the model is only exposed to input images during training (Figure~$3$a). Typically, supervised approaches have demonstrated superior accuracy and specificity to the task and data distribution, but their versatility relates to the availability of paired images, which could be a limitation in live microscopy. Some alternatives~\cite{weigert_content-aware_2018, spahn_deepbacs_2022,xu_cross-modality_2022} propose training models using paired images of \emph{ex vivo} samples—providing perfectly aligned images for training and assessment—to subsequently perform inference with \emph{in vivo} images (Figure~$3$b). Importantly, collecting images from fixed images supports faster creation of more extensive and diverse datasets than live imaging. Depending upon the image features, simulated data may also be viable for training such models~\cite{nehme_deep-storm_2018, saguy-dblinknatmeth-2023, sage-natmeth2019,fang_deep_2021}. However, there remain scenarios where obtaining such paired datasets does not encapsulate the complexity of live experiments, is not experimentally feasible or cross-modality acquisition devices are inaccessible. Given this, alongside time-consuming data annotation processes, has propelled exploration into alternative approaches such as semi- or weakly-supervised~\cite{bilodeau-natmachint-2022}, self-supervised~\cite{krull_noise2void-learning_2019, krull_probabilistic_2020}, or generative techniques~\cite{wang_deep_2019, xu_gan-based_2019, li-commbio-2022} (Figure~$3$a).

Many possibilities exist to exploit deep learning-augmented microscopy and reduce phototoxicity. Drawing inspiration from the delineation provided in~\cite{tian_deep_2021}, we distinguish strategies that aim either to surmount the physical limitations intrinsic to live fluorescence microscopy imaging (\emph{i.e.}, acquisition speed or illumination) or to enhance the content in less qualitatively superior but more sample-friendly image data. The former includes techniques such as denoising, restoration, or temporal interpolation. The latter, referred to by the original authors as "augmentation of microscopy data contrast", includes techniques such as virtual super-resolution~\cite{wang_deep_2019, jin-natcomm-2020, chen_three-dimensional_2021, qiao_evaluation_2021, zhang-opticslasers-2022, qiao_rationalized_2022}.\\

Recent denoising~\cite{chen_deep-learning_2021,krull_noise2void-learning_2019, krull_probabilistic_2020} and restoration approaches \cite{weigert_content-aware_2018, guo_natbiotech_2020, chenj2021, park-natcomm-2022, Li2022, xuesong-natbiotech-2023} have successfully virtualised noise removal, enhanced the SNR, improved fluorescence channel contrast with unparalleled accuracy, or provided isotropic reconstruction of volumetric information of images with low SNR (Figure~$3$c). That is, they support imaging setups with reduced fluorescence illumination that results in lower SNR or reduced optical $3$D sectioning, which in turn, are gentler for the sample. Similarly, acquisition can be slowed down and use intelligent interpolators like CAFI~\cite{Priessner2021} or DBlink~\cite{saguy-dblinknatmeth-2023} to recover temporal information (Figure~$3$d). As we discussed earlier, reducing the number of illumination time points can significantly decrease the cumulative phototoxicity while potentially enabling some degree of photodamage recovery. 

Another innovative approach to enable reduced illumination imaging setups is exploiting cross-modal style transfer methodologies. In brief, these methods involve training a model to translate images between varied microscopy modalities (Figure~$3$e). For example, it was shown that SIM images could be inferred from input images acquired with wide-field illumination~\cite{qiao_evaluation_2021}, reducing the photon dose by a factor of $9$/$15$ in $2$D/$3$D. This capability extends to numerous types of fluorescence microscopy modalities such as confocal to STED~\cite{wang_deep_2019, bouchard_tagan_machineintelligence_2023}, SIM and SRRF~\cite{von_chamier_democratising_2021}, or wide-field to SMLM~\cite{ouyang_deep_2018,nehme_deep-storm_2018, macke_deep_2021}. The enhancement in spatial resolution via learning fine details is similar in objective to traditional deconvolution. It offers comparable benefits for mitigating phototoxicity, such as using less aggressive imaging modalities (widefield and/or confocal against SIM, STED, and SMLM). While the extent these methodologies can contribute towards scientific discovery is up for debate~\cite{belthangady_applications_2019, hutson_artificial_2018}, they undoubtedly elevate image data quality which directly impacts subsequent tasks like accurate tracking or segmentation~\cite{weigert_content-aware_2018, Priessner2021}. As suggested in~\cite{weigert_content-aware_2018}, less aggressive live imaging approaches may yield visually less appealing but easier-to-analyse data due to the reduction in artefacts induced by photodamage (\emph{e.g.}, apoptosis, stressed cells or specimen shrinkage during illumination) with the benefit of preserving close-to physiological conditions.

Exploiting the prowess of data-driven methods, artificial labelling approaches have emerged (Figure~$3$f). They span from inferring cell nuclei (\emph{e.g.}, Hoechst staining) from actin (e.g., Lifeact staining)~\cite{von_chamier_democratising_2021}, to estimating specific fluorescence information (\emph{e.g.}, nucleoli, cell membrane, nuclear envelope, mitochondria or neuron-specific tubulin) from brightfield input images \cite{ounkomol_label-free_2018, christiansen_silico_2018}. It's worth noting that the latter technique is also categorized as a cross-modality style transfer approach. Moreover, artificial labelling can also be employed for channel unmixing~\cite{mcrae_robust_2019,xue-bioinformatics-2022, jiang-biorxiv-2023}, which offers several key advantages, including illumination channel reduction, acceleration of image acquisition, or support for more straightforward or cost-effective imaging experiments (Figure~$3$f). Among these benefits, the former is pivotal in enabling more sample-friendly setups. Indeed, artificial labelling often works as an intermediary step for further quantification such as segmentation or tracking~\cite{hollandi_nucleaizer_2020, von_chamier_democratising_2021}.

Beyond the questions present also in the natural domain, novel inpainting and modality transfer techniques pose an additional challenge in bioimaging: they must accurately infer reliable and quantifiable physiological information. 
Thus, further validation, more assessment methodologies, and standard quantitative strategies ought to ascertain both the biological reliability of restored images and the integrity of recovered signal intensity (\emph{i.e.,} virtual images) are needed~\cite{belthangady_applications_2019, laine-natmethods2021, lambert-natmethods-2023}.

Biomedical data typically exhibit high variability due to factors such as the biological sample's physiology, experimental protocol, imaging setup, and even the individual researcher conducting the experiments. Defining and establishing an accurate ground truth becomes crucial~\cite{laine-natmethods2021, maska-natmeth-2023}, as deep learning model training highly depends on the data quality. Quality in terms of image processing ranges from the number of images available to train to how suitable the features of these images are for the task to train the model on. For instance, it remains to be seen how to strike an optimal balance between data acquisition and model accuracy (\emph{i.e.}, defining an ideal sampling frequency for precise cell tracking). A prevalent notion suggests that larger training datasets enhance model performance. Simultaneously, many publicly available annotated or paired databases are growing~\cite{ caicedo_nucleus_2019, ouyang-hpa-2019, maska-natmeth-2023}. However, strategies to combine these datasets effectively while preserving each experimental setup's specificity still need to be clarified. Furthermore, pre-trained models that facilitate fine-tuning and transfer learning are readily accessible thanks to trained deep learning model repositories such as the Bioimage Model Zoo~\cite{, ouyang_bmz_2022} and MONAI~\cite{cardoso-monai-2022}. Nonetheless, further enquiry is necessary to ascertain the best practices for assembling adequate training data and executing effective transfer learning while considering data quality, image features, and task-specific analytical requirements. Because live-cell image data is usually highly redundant, such optimisation could maximise information extraction from images while minimising photodamage during the acquisition. 

Life scientists hold the top authority in deciding parameters such as sampling frequencies, resolution, or fields of view. While these decisions may sometimes be sub-optimal for subsequent quantification, they are our best reference for accurate performance. Incorporating users in the loop or their expertise as priors to guide model performance towards more specific and biologically relevant results is one of the pro missing directions to take into AI-enhanced microscopy. Recent technological advances such as the analytical representation of sparse and raw information to create priors - an approach already proposed for the segmentation of natural pictures in works such as the Segment Anything Model (SAM)~\cite{kirillov-sam-2023} - is an important step towards it.

\section*{Future Outlook}
\begin{figure*}
\centering
\includegraphics[width=1\linewidth]{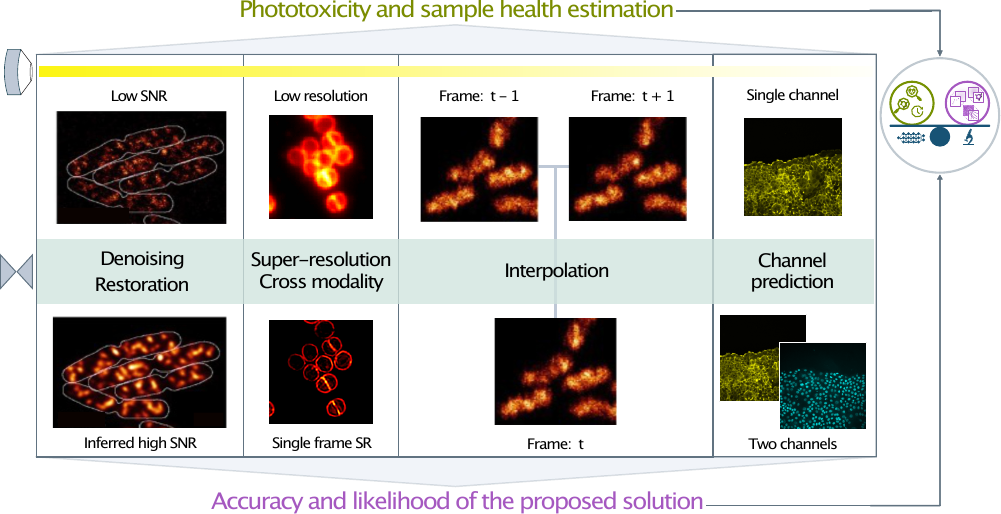}
\caption{\textbf{ AI-enhanced photodamage-aware microscopy}. In practice, a more gentle imaging setup would balance the health state of the imaged sample and the quality of the information obtained with deep learning augmented microscopy. Several aspects, such as the levels of phototoxicity, the properties of the extracted image information, and the different tasks that one could perform with deep learning, their expected accuracy and their likelihood to fit well in the experiment must be considered when deciding an optimal imaging configuration. In this way, combining advanced image processing in microscopy acquisition pipelines would support reproducible and optimised conditions for live-cell imaging. The source images for this figure are provided by~\cite{von_chamier_democratising_2021, spahn_deepbacs_2022}}
\label{fig:figure4}
\end{figure*}

With the advance of fluorescence microscopy technology, the field is developing intelligent imaging techniques to minimise photodamage and enable accurate observations of biological dynamics. Similar to automatic cars or smart robots in industrial environments, microscopes can be empowered with AI components that enable real-time decisions by integrating the information extracted from the observed data into an intelligent feedback loop that balances the health of the sample and the image data quality (Figures~$2$c and~$4$) ~\cite{Scherf_gentlemic_2015}. Note that we refer to image quality as the combination of features (\emph{e.g.,} resolution in time and space, SNR, the size of the field of view, or the number of fluorescent channels) that allow getting the most information relevant for the understanding of the observed process.   In line with these, some have already conceptualised and proved such systems. In the first place, we find event-driven approaches, which automatically identify specific objects or incidents in the image that trigger the acquisition in real-time~\cite{alvelid-natmeth-2022, zachar-natcomm-2022, Mahecic2022, chiron_scirep_2022, andre-cellreports-2023}. While these adaptive approaches reduce the induced phototoxicity by only illuminating the sample when needed, in most cases, they are equipped with deep learning models trained to recognise predefined objects or elements in the images, which is not always the case in biology and may limit, even bias, the observation of novel physiological processes. Alternative approaches propose the integration of image resolution enhancement in the loop to get faster and gentler setups. In~\cite{wagner-natmeth-2021}, a deep learning model is trained and validated in the acquisition loop to enhance the volumetric reconstruction and provide an adaptive light field microscopy (LFM) setup. In the context of super-resolution imaging, Bouchard \emph{et al.}~\cite{bouchard_tagan_machineintelligence_2023} propose evaluating the quality of virtually inferred STED images from confocal microscopy images to determine the uncertainty in the observed sample and determine whether a new STED image should be acquired or not. All these works pose new paradigms in the realm of smart microscopy.

Deep learning approaches, particularly the unsupervised ones, learn and match data distributions even in highly heterogeneous or complex scenarios without the need for human descriptions or annotations. As mentioned in~\cite{pinkard-waller-2022}, such methods could be exploited to identify the events that deviate from the general distribution, \emph{i.e.}, to discover new biological patterns. Thus, advancing generative models and unsupervised/self-supervised approaches that can effectively learn from unpaired data alone can provide flexibility when paired datasets are difficult to obtain experimentally or when investigating new dynamics, and contribute to unbiased observations.

Straightforwardly, one would realise that despite sample health preservation being a strong motivation and a major limitation in live-cell imaging, none of the existing solutions analyses or estimates it directly. Robust photodamage reporters that provide quantitative assessments of sample health without requiring additional fluorescence channels can directly contribute to more reproducible biological readouts. This could involve exploring modalities like transmitted light microscopy or label-free techniques. Namely, quantitative reporters would support the design of automated workflows that analyse sample health in real-time during image acquisition rather than only evaluating the image quality (Figure~$4$). This would allow the detection of early signs of photodamage and the adaptive determination of optimal imaging conditions. In other words, it will open the door for data-driven sample-oriented live microscopy. 

Pursuing such technical innovations while deepening our understanding of photodamage mechanisms will enable microscopists to unlock the full potential of intelligent imaging. With photodamage-aware AI and automated tools, the goal of observing undisturbed physiological processes can be realised. This will profoundly enhance fluorescence microscopy's capacity to uncover ground truths in biology.

\section*{Discussion}
Fluorescence microscopy has become an indispensable tool in cell biology, providing unparalleled insights into biomolecular dynamics. However, phototoxicity remains a major impediment, necessitating a deeper mechanistic understanding alongside the development of imaging techniques that mitigate this limitation. While the intercourse between microscopy hardware innovation and computational imaging has yielded promising solutions, standardised methodologies to comprehensively assess photodamage are lacking. Recent strides in deep learning provide optimism by enhancing information extraction from low-light or accelerated acquisitions. Nonetheless, robust validation strategies are still required to ensure biological fidelity.

A key opportunity lies in constructing universal photodamage metrics that account for subtle, cumulative deviations in sample physiology. Integrating such quantifications in intelligent automated analysis enables microscopes to optimise imaging conditions dynamically. This calls for a convergence of synergistic advancements spanning photodamage biology, microscopy hardware, computational imaging and model interpretation.
An outstanding challenge is model training, which requires extensive paired datasets that sufficiently encapsulate biological variability. Alternatives like unsupervised learning provide flexibility but may compromise accuracy. Incorporating biological expertise through techniques like priors and prompts appears promising to guide models. Additionally, optimised strategies for effective model training and validation must be developed through empirical examination.

While deep learning has elevated imaging capabilities, over-reliance on its reconstruction prowess could promote complacency. Achieving gentler acquisition first requires re-evaluating illumination intensities and sampling frequencies. We claim that AI should extract the maximum information from the least invasive data rather than recover information already compromised by excessive phototoxicity. Keeping biological relevance at the crux while exploiting technology will lead to microscopes that truly observe life undisturbed.\\

\begin{contributions}
E.G.M., M.DR., and R.H. conceptualised the majority of the manuscript. E.G.M. and M.DR. wrote the manuscript with input from J.W.P., G.J., and R.H.  J.W.P. and G.J. contributed text, figure design, critical comments, and conceptual suggestions to improve the manuscript. J.W.P., G.J., and R.H. reviewed and edited the manuscript.
\end{contributions}

\begin{acknowledgements}
M.DR., E.G.M. and R.H. acknowledge the support of the Gulbenkian Foundation (Fundação Calouste Gulbenkian), the European Research Council (ERC) under the European Union’s Horizon 2020 research and innovation programme (grant agreement No. 101001332), the European Commission through the Horizon Europe program (AI4LIFE project with grant agreement 101057970-AI4LIFE, and RT-SuperES project with grant agreement  101099654-RT-SuperES), the European Molecular Biology Organization (EMBO) Installation Grant (EMBO-2020-IG-4734) and the Chan Zuckerberg Initiative Visual Proteomics Grant (vpi-0000000044 with DOI:10.37921/743590vtudfp). E.G.M. was also supported by EMBO Postdoctoral Fellowship (EMBO ALTF 174-2022). This study was supported by the Academy of Finland (338537 to G.J.), the Sigrid Juselius Foundation (to G.J.), the Cancer Society of Finland (Syöpäjärjestöt; to G.J.), and the Solutions for Health strategic funding to Åbo Akademi University (to G.J.). This research was supported by InFLAMES Flagship Programme of the Academy of Finland (decision number: 337531).
\end{acknowledgements}

\section*{Bibliography}

\end{document}